\documentclass[referee]{raa}            % referee version: for submission

\usepackage{graphicx,times}             %for PS/EPS graphics inclusion, new
\graphicspath{{fig/}}
\usepackage{cite}
\usepackage{natbib}
\usepackage{color}
\usepackage{amssymb,amsmath}
\usepackage{hyperref}

\hypersetup{pdftitle =paper, pdfauthor =G Li, pdfsubject=SEPs, pdfkeywords = SEP CME }
\hypersetup{colorlinks = true, linkcolor = blue, anchorcolor = red, citecolor = blue, filecolor = red, urlcolor = blue}

\newcommand{\beq}{\begin{equation}}
\newcommand{\eeq}{\end{equation}}
\newcommand{\beqa}{\begin{eqnarray}}
\newcommand{\eeqa}{\end{eqnarray}}

\begin{document}
\bibliographystyle{raa}
\title{Modeling the 10 September 2017 solar energetic particle event using the iPATH model}

   \volnopage{Vol.0 (200x) No.0, 000--000}      %%preserved for Editor. DOn't remove!
   \setcounter{page}{1}          %%starting page, preserved for Editor. DOn't remove!

\author{Zheyi Ding \inst{1,2}, Gang Li\inst{2,*},
Junxiang Hu\inst{2}, Shuai Fu\inst{2,3}}

\institute{
{School of Geophysics and Information Technology, China University of Geosciences (Beijing), Beijing 100083, China; } \\
{Department of Space Science and CSPAR, University of Alabama in Huntsville, Huntsville, AL 35899, USA;\it gangli.uah@gmail.com}\\
{Lunar and Planetary Science Laboratory, Macau University of Science and Technology, Macau 519020, China}\\ }

   \date{Received~~2019 month day; accepted~~2019~~month day}

%% \linenumbers*[1]

\abstract{On September 10 2017, a fast coronal mass ejection (CME) erupted from the active region (AR) 12673, leading to a ground level enhancement (GLE) event at Earth.
Using the 2D improved Particle Acceleration and Transport in the Heliosphere (iPATH) model, we model the large solar energetic particle (SEP) event of 10 September 2017 observed at Earth, Mars and STEREO-A.
Based on observational evidence,we assume that the CME-driven shock experienced a large lateral expansion shortly after the eruption, which is modelled by a double Gaussian velocity profile in this simulation. We use the in-situ shock arrival times and the observed CME speeds at multiple spacecraft near Earth and Mars as constraints to adjust the input model parameters.
The modelled time intensity profiles and fluence for energetic protons are then compared with observations. Reasonable agreements with
 observations at Mars and STEREO-A are found. The simulated results at Earth differ from observations of GOES-15. Instead, the simulated results at a heliocentric longitude 20$^\circ$ west to Earth fit reasonably well with the GOES observation. This can be explained if the pre-event solar wind magnetic field at Earth is not described by a nominal Parker field. Our results suggest that a large lateral expansion of the CME-driven shock and a distorted  interplanetary magnetic field due to previous events can be important in understanding this GLE event.
}

   \authorrunning{Z.-Y. Ding et al. }            %author_head in even pages
   \titlerunning{Modeling solar energetic particle event of 10 September 2017}         % title_head in odd pages

   \maketitle
%% The author head (on even pages) and the title head (on odd pages) will be
%% automatically extracted from \author{} and \title{}. Whenever the title is too long,
%% you will be asked to supply a shorter one by inserting either \authorrunning{} or
%% \titlerunning{} before \maketitle. Anyway, you can specify your own heads.
%%
%%
%% Note: In the following text body of your manuscript, please note several differences from
%%       other major journals:
%% (1) \subsection{Please Capitalize the First Letter of Each Notional Word in Subsection Title}
%% (2) Please Capitalize the First Letter of Each Notional Word in all tables' captions

%
%________________________________________________ sections below
%
\section{Introduction}           %% first-level sections will be auto-capit  alized
\label{sec.intro}

Solar energetic particle events are historically  classified into two broad categories: impulsive and gradual events  (\citealt{Cane+etal+1986,Reames+1995,Reames+1999}). In this paradigm, impulsive SEPs are accelerated at solar flares and propagate along the interplanetary magnetic field  to Earth with a rapid rise and decay phase in the particle time intensity  profiles. In comparison, gradual SEPs are accelerated via the diffusive shock acceleration mechanism at CME-driven shocks. Because shocks are of large scale and last much longer than flares, these events are characterized by a prolonged intensity profile and often higher fluences than impulsive events. Large gradual SEP events are of particular concern because the accompanying high-energy protons pose the most serious radiation threats to astronauts living beyond low-Earth orbit and can damage electronics on satellites in the space (\citealt{Desai+Giacalone+2016}). In some largest events, accelerated particles can reach energies up to several GeV, leading to significant increases in particle count rates through neutron monitors at the Earth's surface and are known as Ground Level Enhancements. Large SEP events are usually associated with fast shocks and high ambient energetic particle intensity (referred as seed particles) prior to the shock  (\citealt{Kahler1996,Kahler+etal+2000}). The increasing abundance of the seed particles is possibly  from preceding flares   (\citealt{Mason+etal+1999}) or preceding CMEs  (\citealt{Gopalswamy+etal+2004,Li+zank+2005b}). \citet{Li+etal+2012} proposed a ¡°twin-CME¡± scenario for GLE events in which two CMEs erupt from same or near active region within a period of 9 hours. The preceding CME and its driven shock  disturbs the coronal and interplanetary environment and leads to enhanced turbulence level which facilitates a more  efficient acceleration at the second CME-driven shock. The scenario  was later extended to large SEP events (\citealt{Ding+etal+2013}).

The most recent GLE event occurred on September 10 2017, classified as GLE 72 (\citealt{Mishev+etal+2018}). A X8.2 class solar flare erupted around 15:35 UT from AR 12673 at S09W88, followed by a wide and very fast Coronal Mass Ejection. The eruption was well observed by multiple spacecraft at different longitudes, provided a stereo view of this extreme case.   The halo CME was first observed by the Solar and Heliospheric Observatory (SOHO) Large Angle and Spectrometric Coronagraph (LASCO) at 16:00 UT, with a reported linear speed of $\sim$3136 km/s from the CDAW catalog. Before this extreme eruption, there were multiple preceding eruptions between 4 September and 9 September and the eruption directions varied from the longitude of 4$^\circ$ to 105$^\circ$ (\citealt{Luhmann+etal+2018}). However, this event would not be classified as a "twin-CME" (\citealt{Li+etal+2012}), because the closest preceding CME erupted on 23:12 UT on 9 September 2017,  $\sim$17 hours before the main eruption, exceeding the time interval threshold of $13$ hours (\citealt{Ding+etal+2014}) as a twin-CME.  The global EUV waves (\citealt{Liu+etal+2018,Hu+etal+2019}),  signatures of a flux rope and a long current sheet (\citealt{Seaton+etal+2018,Warren+etal+2018,Yan+2018}) were observed by the Solar Dynamics Observatory (SDO).
 \citet{Liu+etal+2019} studied the geometry and  kinematics of the CME-driven shock.
They argued that the close shock arrival times at Earth and Mars is a result of large lateral expansion of the shock early in the eruption.
They suggested that a large lateral expansion of the shock can affect particle acceleration and consequently the observations at different longitudes. Large lateral expansion of the CME-driven shock for large CME eruptions have been investigated previously.  \citet{Gopalswamy+etal+2012} examined the white-light CME evolution  and noted a rapid and large lateral expansion of the CME erupted on 2010 June 13.  \citet{Kwon+etal+2015} noted the apparent width of halo CMEs as seen from multiple spacecraft was related to the expanding shock with a 360$^\circ$ envelope. Recent works (e.g. \citealt{Liu+etal+2017,Liu+etal+2019, Zhu+2018})
have suggested that lateral expansions of shock occur frequently in large big eruptions. However, the effects of lateral expansion on solar energetic particles have not been considered before. In this work, we will examine this problem.

Energetic particles associated with this fast CME were observed at Earth and Mars as well as Solar Terrestrial Relations Observatory A  (STEREO-A). GLE was observed by several neutron monitors stations at about 16:15 UT on 10 September 2017  (\citealt{Mishev+etal+2018,Guo+etal+2018}). The different energy channels from the instruments on board Geostationary Operational Environmental Satellite 15 (GOES-15) show an intensive and long-lasting enhancement of energetic particles intensity (\citealt{Guo+etal+2018,Bruno+etal+2019}). This event observed at Earth has a soft spectrum at high energies compared with most of GLE events in the solar cycle 23  (\citealt{Gopalswamy+etal+(2018),Cohen+Mewaldt+2018,Bruno+etal+2019}). On 10 September at around 19:50 UT, the high energy protons were detected by Radiation Assessment Detector (RAD)  (\citealt{Hassler+etal+2012}) from Mars Science Laboratory (MSL) (\citealt{Zeitlin+etal+2018,Ehresmann+etal+2018,Lee+etal+2018,Guo+etal+2018}). Later, high energetic protons were observed with a slow increase at STEREO-A at around 08:00 UT on 11 September 2017 (\citealt{Guo+etal+2018,Lee+etal+2018}).

Attempts of modeling particle acceleration and transport of SEP events have been undertaken by many groups  (e.g. \citealt{Heras+etal+1995,Kallenrode+1997,Luhmann+etal+2007,Luhmann+etal+2010,Luhmann+etal+2017}). These models treat CME-driven shock as a source of energetic particles without acceleration process. A dynamic onion-shell model of the strong shock propagation and particle acceleration has been developed by \citet{Zank+etal+2000}. This model was improved by \citet{Rice+etal+2003} for shocks with arbitrary intensities, and by \citet{Li+etal+2003} to model the transport of energetic particles using a Monte-Carlo approach. \citet{Li+etal+2005a} further extended the model to include heavy ions. A comprehensive numerical model, developed by these authors, called Particle Acceleration and Transport in the Heliosphere(PATH) model.  Modeling specific SEP events using the PATH model shows their reasonable agreement  (\citealt{Verkhoglyadova+etal+2010,Verkhoglyadova+etal+2009}). Recently, \citet{Hu+etal+2017} have extended the PATH to a 2D model, named improved Particle Acceleration and Transport in the Heliosphere (iPATH). The new model has the capability to study the  characters of particle acceleration and transport at a 2D shock in multiple locations of the ecliptic plane. \citet{Hu+etal+2018} modelled an example gradual SEP event as observed at multiple locations.

In the iPATH model, the shock speed profile at the inner boundary is assumed to be a Gaussian form in longitude and the propagation direction of the shock is assumed to be radial. Such an assumption is for simplicity.  As we discussed earlier, large lateral expansion of shocks can be common for large events. A lateral expansion implies that the open angle of the shock will increase over time. This means that the treatment of the inner boundary of the shock profile in the iPATH model needs to be improved. In this work, we modify the inner boundary of the shock profile to examine the effect of the large lateral expansion on producing solar energetic particles. We note that because the background solar wind is unlikely to be homogeneous, the lateral expansion and the shock profiles do not need to be homogeneous. Therefore there is no need to assume a symmetric eruption.

In this paper, we discuss our modeling of the 10 September 2017 GLE event using the iPATH model. In section \ref{sec.model}, we describe the iPATH model and the model setup.  In section \ref{sec.result}, we discuss the CME-driven shock configuration and shock parameters. Modelled time intensity profile are compared with GOES-15 observation results in different energy channels. We integrate the data of proton flux from the Advanced Composition Explorer (ACE) and GOES-15 and fit the spectrum using the Band function. The modelled spectrum is compared with observation fitted results. We also compare the modelled time intensity profiles with the observation results at Mars and STEREO-A. We then  discuss our results and conclude in section \ref{sec.Conc}.

\section{Model}
\label{sec.model}

The iPATH model is a two-dimensional MHD code plus a particle transport code developed by \citet{Hu+etal+2017}.  It is a continuation of the earlier 1D PATH model  (\citealt{Zank+etal+2000};\citealt{Li+etal+2003,Li+etal+2005a}), addressing solar energetic particle events in the ecliptic plane. It contains two separate modules: the first module models the background solar wind and follow particle acceleration at the shock front, the second module models the transport of SEPs escaped from the upstream of the CME-driven shock. Here we limit ourselves to the modifications that we introduce to the iPATH in modeling the 10 September 2017 event.

We model the propagation of background solar wind and CME-driven shock limited only in the ecliptic plane using a 2D MHD code. For simplicity, the background solar wind is assumed to be homogeneous.  We use the $8$-hour averaged in-situ solar wind observation near Earth before the CME eruption as the solar wind inputs. The CME-driven shock is treated by perturbing the inner boundary  (proton number density $n$, solar wind speed $V_{sw}$ and temperature $T$) at 0.05 AU for a short period of time (e.g. 1 hour). Such a simplified treatment is similar to  the WSA-ENLIL+Cone model for the space weather prediction.  The inner boundary is set at $0.05$ AU ($\sim10R_{s}$) in this work,  thus we can not model particle acceleration in the low corona. Modeling CME eruption in the low corona needs a more detailed description about the magnetic field and the corona condition. However, CME-driven shock can  be formed in the low corona (\citealt{Gopalswamy+etal+2013}, \citealt{Liu+etal+2009}) and some highest energy particles are produced within several solar radii. Modeling this part of SEPs is beyond the scope of the iPATH model.  In this work, we introduce two Gaussian velocity profiles to account for both a lateral expansion of the shock and any inhomogeneity in the background solar wind.
Specifically, we consider the following
eruption profile:

\begin{equation}
A(\phi)=\left\{
\begin{array}{lcl}
A_{1} e^{-\frac{(\phi-\phi_{c_{1}})^{2}}{2 \sigma_{1}^2}}       &      & {0  < t < D_{1}}\\
A_{2} e^{-\frac{(\phi-\phi_{c_{2}})^{2}}{2 \sigma_{2}^2}}H[\phi-\phi_{min}]*H[\phi_{max}-\phi]        &      & {0  < t-t_{s} < D_{2}}\\
\end{array} \right.
 \label{eq:guass}
\end{equation}
where $A_{1},A_{2}$ are the perturbed model parameters (number density, speed, temperature) at longitudes $\phi_{c_{1}}$ and  $\phi_{c_{2}}$. The angles $\phi_{c_{1}}$ and  $\phi_{c_{2}}$ are the central longitude of the two Gaussian distributions. The variances $\sigma_{1},\sigma_{2}$ are related to the full width at half maximum (FWHM) of the perturbed width. For the second profile, we identify a range given by $[\phi_{min},\phi_{max}]$ as the perturbed range associated with the lateral expansion. $D_{1},D_{2}$ are the perturbing duration of two eruption profiles, and $t_s$ is the start time of the second eruption, which we take as the start time of  the lateral expansion. $H$ is the Heaviside function. For simplicity, the background interplanetary magnetic field is assumed to be a Parker spiral,

\begin{equation}
B_{r}=B_{0}\left (\frac{R_{0}}{r} \right )^{2};B_{\phi}=B_{r}\left (\frac{\Omega r\sin{\theta}}{u_{sw}}\right) \left(r \gg R_{0} \right),
\label{eq:b}
\end{equation}
where $B_{r}$ and $B_{\phi}$ are the radial and azimuthal component of the IMF at  heliocentric distance $r$. $u_{sw}$ is solar wind speed. $\theta=90^{\circ}$ corresponds to the magnetic field in the ecliptic plane. $B_{0}$ is the radial  component of the IMF at $R_{0}$.

After perturbing the inner boundary,  the CME-driven shock is followed in the code and the shock parameters are calculated at every time step. From these shock parameters we obtain the dynamic timescale $t_{dyn} = \frac{R}{dR/dt}$. Balancing $t_{dyn}$ with the acceleration time scale yields the maximum particle momentum  $p_{max}$ (\citealt{Drury+1983}):

\begin{equation}
t_{dyn} = \int_{p_{0}}^{p_{max}}\frac{3s}{s-1}\frac{\kappa}{U_{shk}^{2}}\frac{1}{p}dp,
\label{eq:dynamic time}
\end{equation}
where $p_{0}$ is the injection momentum, $s$ is the shock compression ratio, $\kappa$ is the diffusion coefficient of particle, $U_{shk}$ is the shock speed in the upstream frame.
Once $p_{max}$ is obtained, the particle distribution function in the outmost parcel$(j_r,k_\phi)$ at each time step $t_k$ is given by,

\begin{equation}
f(j,k,p,t_k) = c_1*\epsilon_{j,k}n_{t_k,k}p^{-\beta}H[p-p_{inj}^{j,k}]*H[p_{max}^{j,k}-p]\exp\left(-\frac{E}{E_{0}}\right)
\label{eq:fp}
\end{equation}
where $\beta = \frac{3s_{j,k}}{s_{j,k}-1}$, $\epsilon_{j,k}$ is the injection efficiency, $n_{t_k,k}$ is the upstream solar wind density at the time $t_k$ in front of the parcel $(j_r,k_\phi)$. $p_{inj}$ is the particle injection momentum. $E$ is the particle energy and $E_{0}$ is the kinetic energy corresponds to the maximum particle momentum $p_{max}$ obtained in the equation~(\ref{eq:dynamic time}).  In the above $H$ is the Heaviside function. $c_1$ is a normalization constant,
\begin{equation}
c_1 = 1/\int_{p_{inj}^{j,k}}^{p_{max}^{j,k}}p^{-\beta}\left \{H[p-p_{inj}^{j,k}]*H[p_{max}^{j,k}-p]\right \}d^3p
\end{equation}
Note that this functional form is different from previous works  (e.g. \citealt{Li+etal+2005a,Hu+etal+2017}), where the exponential $exp(-E/E_0)$ tail was not included.
In \citet{Drury+1983}, the dynamical time
$t_{dyn}$ in equation~(\ref{eq:dynamic time}) is to be understood as the average
time for accelerating particles from $p_{inj}$ to $p_{max}$. Clearly, some portion of these particles would take longer time than $t_{dyn}$. However, when $t>t_{dyn}$, shock parameters can not be regarded as constant anymore, so we expect a roll-over of $f(p)$ near $p_{max}$. This, of course, is the consequence of a finite acceleration time. In an earlier paper, \citet{Forman+Drury1983} examined
the effect of finite time analytically and showed that the particle distribution function at higher momentum is of exponential decay of the diffusion coefficient, which is a function of particle momentum.  \citet{Ellison+Ramaty+1984} adopted an exponential decay tail at high energy $\sim exp(-E/E_0)$ to account for the effect of finite shock size and finite acceleration time.
Some recent numerical simulations by  \citet{Zuo+etal+2011} and \citet{ Kong+etal+2019} which examined particle acceleration at a prescribed shock showed that
the time dependent particle spectra are consistent with a power law with an exponential tail. With these considerations,
we adopt equation~(\ref{eq:fp}).

To model SEP events at multiple locations simultaneously, the iPATH model uses
a 2D onion shell module to keep track energetic particles in the shock complex.  At the $j$-th time step, the $j$-th shell (the outermost shell) is generated and divided longitudinally into different parcels $(j_r,k_\phi)$ with an angular width of $5^\circ$.  Particles accelerated at the shock front experience convection with the parcel and diffusion between parcels. Note that particles only diffuse in parcels whose $p_{max}^{j,k}$ are greater than particles' momentum $p$. This is because for parcels with $p_{max}^{j,k} < p$,  there is no excited wave  turbulence that can trap these particles.
Particles can escape when they diffuse far enough ahead of the shock. \citet{Zank+etal+2000} assumed an escape length $l = 4 \lambda_{esc}$, where $\lambda_{esc} = \frac{\kappa_{rr}}{U_{shk}}$ is the scatter length scale and $\kappa_{rr}$ is the particle upstream diffusion coefficient in the direction of shock normal. Within the length $l$, the excited wave density is significantly higher than that of the ambient solar wind. If the wave intensity has no $x$-dependence, then the particle distribution function decay exponentially with the escape length in the shock front. This allows one to obtain the escaped particle distribution function. %Beyond the escape length, particles escape into the unperturbed solar wind.
Alternatively \citet{Li+etal+2005a} calculated the escaped particle number explicitly instead of using the particle distribution function. Note, the escape boundary is related to each individual parcel.  %maximum particle momentum $p_{max}^{j,\phi}$.
In this work, we follow \citet{Li+etal+2005a} and obtain the escaped particle number of momentum $p$ at time $t_{k}$ and at longitude $\phi$ related to the outmost parcel $(j_r,k_\phi)$ by,

\begin{equation}
N_{esc} (k,p,t_{k}) = \sum_{j=1}^{J(k,p)} \frac{1}{2\sqrt{\pi}}\frac{\Delta R^{\ast}N_{j,k}\left ( t_{k}-1 \right ) }{r_{j,k}^{\ast}}\left \{ \exp(-x^{2}) + \frac{\sqrt{\pi }r_{j,k}^{\ast}}{\Delta R^{\ast}}\left [ 1-erf(x) \right ]\right \},
\label{eq:escape number}
\end{equation}
where $N_{esc} (k,p,t_{k})$ is the number of particles  escaped from shock complex at time $t_{k}$ and longitude $\phi$ within $d^{3}p$ phase space;
$ J(k,p)$ is the shell number for which $p_{max}=p$;  $(\Delta R^{\ast})^{2} = 4\kappa_{j,k}*(t_{k}-t_{k-1})+2([r_{j+1,k}(t_{k})-r_{j,k}(t_{k})]/2)^{2}$ decides how many particles escape from the parcel $(j,k)$. This parameter is larger for high energy particles since the diffusion coefficient $\kappa$ increases with momentum.  $N_{j,k}\left ( t_{k}-1 \right )$ is the particle number in the parcel ($j,k$) before diffusion, $r_{j}^{\ast}=(r_{j,k}+r_{j+1,k})/2$ represents the heliocentric distance of the center of parcel $(j,k)$, $x=(r_{esc}-r_{j,k}^{\ast})/\Delta R^{\ast}$ is a dimensionless parameter, and $r_{esc} = r_{J_{(k,p)},k}+l$ represents the escaping boundary for particles of momentum $p$.

Once particles escape from the shock complex, they propagate along the IMF in the relatively undisturbed solar wind.  We follow closely \citet{Hu+etal+2017} in describing the transport of escaping  particles in the second module of the iPATH model. In particular, we also include the cross-field diffusion and use the backward stochastic differential equation method.  The cross-field diffusion is governed by a term $ \sim \triangledown \cdot \left (\kappa_{\perp} \cdot \triangledown f \right )$ (see equation (18) of \citet{Hu+etal+2017}), where  $\kappa_{\perp}$ is the perpendicular diffusion coefficient of energetic particles,  \citet{Hu+etal+2017}  calculated $\kappa_{\perp}$ from the parallel diffusion coefficient $\kappa_{\parallel}$ based on the Non-Linear Guiding Center theory (NLGC).
In the NLGC theory, the solar wind turbulence is assumed to have a geometry of "slab+2D", where the slab component describes Alfv\'{e}nic fluctuation along the background field and the 2D turbulence describes how the background field line (here the Parker field) meanders.  An important parameter describing the 2D component of the turbulence is the bend-over scale  $l_{2D}$ (\citealt{Shalchi+etal+2010}). In the work of \citet{Hu+etal+2017}
the radial dependence of  $l_{slab}$ and $l_{2D}$ was ignored.  \cite{Hu+etal+2018} considered a more general case of,
\beq
  \begin{split}
l_{slab} \sim r^{\alpha};l_{2D} \sim r^{\alpha},
  \end{split}
\label{eq:l2d}
\eeq
where $\alpha$ is a parameter and close to $1$. In this work, we set $\alpha = 0.8$ and assume the turbulent magnetic field square $\delta B^{2} \sim r^{-3}$, thus the radial dependence of $\kappa_{\parallel}$ and  $\kappa_{\perp}$ used in \cite{Hu+etal+2018} has the following form:
\beq
  \begin{split}
%\frac{\kappa _{\perp }}{\kappa _{\parallel }} \sim v^{-2/9}B^{-22/9}r^{-29/9}.
\kappa _{\parallel} \sim v^{\frac{4}{3}}B^{\frac{5}{3}}r^{\frac{2}{3}\alpha+3};
\kappa _{ \perp } \sim v^{\frac{10}{9}}B^{-\frac{7}{9}}r^{\frac{8}{9}\alpha-1}
  \end{split}
\label{eq:kappa}
\eeq
where $v$ is particle speed, $B$ is the background magnetic field, and $r$ the heliocentric distance. In this work, we set ${\kappa _{\perp }}/{\kappa _{\parallel }}$ to be $0.0099$ at 1AU for 1 MeV proton. With this choice of $\kappa_{\perp}$, we examine cross field transport of energetic particles and discuss the observations at Mars and STA, which, as we show below, depend strongly on
 $\kappa_{\perp}$.

\section{Results}
\label{sec.result}
For our simulation, we average $8$ hours of solar wind observation before the CME eruption as a guide to the background solar wind inputs. This yields a solar wind speed as $500$ km s$^{-1}$, a magnetic field of $3.9$ nT and a number density of 0.8 cm$^{-3}$  at 1 AU. Note that there were some preceding CMEs, which mostly affects the number density of these three parameters. To account for the recovery of the corona after the preceding CME, we use the 8-hours averaged number density of 3.5 cm$^{-3}$ before shock arrival. As shown in equation~(\ref{eq:guass}), we use a double Gaussian profile for the velocity disturbance and the corresponding parameters describing these two Gaussian profiles are listed in Table~\ref{table1}. To model the large lateral expansion, we use a large variance for the second Gaussian distribution but limit the perturbation only within the longitudes of  $[\phi_{min},\phi_{max}]=[10,90]$. The perturbation duration ($D_{1},D_{2}$) both are set to be $1$ hour and the start time ($t_s$) for the second perturbation is at $1$ hour. The injection efficiency is assumed to be 1$\%$ for $\theta_{BN}=0$ (i.e. parallel shock geometry). We set the ambient turbulence level $\delta B^{2}/B^{2}$ to be $0.5$ at 1 AU.

\begin{table}

\bc
\begin{minipage}[]{100mm}
\caption[]{The initial CME parameters of double Gaussian distributions \label{table1}}\end{minipage}
\setlength{\tabcolsep}{2.5pt}
\small
 \begin{tabular}{ccccccc}%{ p{2cm}p{2cm}p{2cm}p{2cm}p{1cm}}
  \hline\noalign{\smallskip}
 Gaussian distribution  &    Initial speed    &   Density amplitude  & Perturbation center   &   variances     \\$i$&km s$^{-1}$& &$\phi_{c_{i}}$&$\sigma_{i}$\\
  \hline\noalign{\smallskip}
1 & 2200 & 4& 115$^{\circ}$ &44$^{\circ}$ \\
2 & 1400 & 3.4& 90$^{\circ}$ &120$^{\circ}$ \\
  \noalign{\smallskip}\hline
\end{tabular}
\ec

%% place \tablecomments and \tablerefs below \end{center| and \end{center}:
%% you may leave the table-width parameter to editors or set to your actual size
\tablecomments{0.86\textwidth}{
Density amplitude refers to the ratio of the enhanced number density to that of the background at the inner boundary.}
\end{table}

\begin{figure}[htb]
    \centering
    \noindent \includegraphics[width=1.0\textwidth]{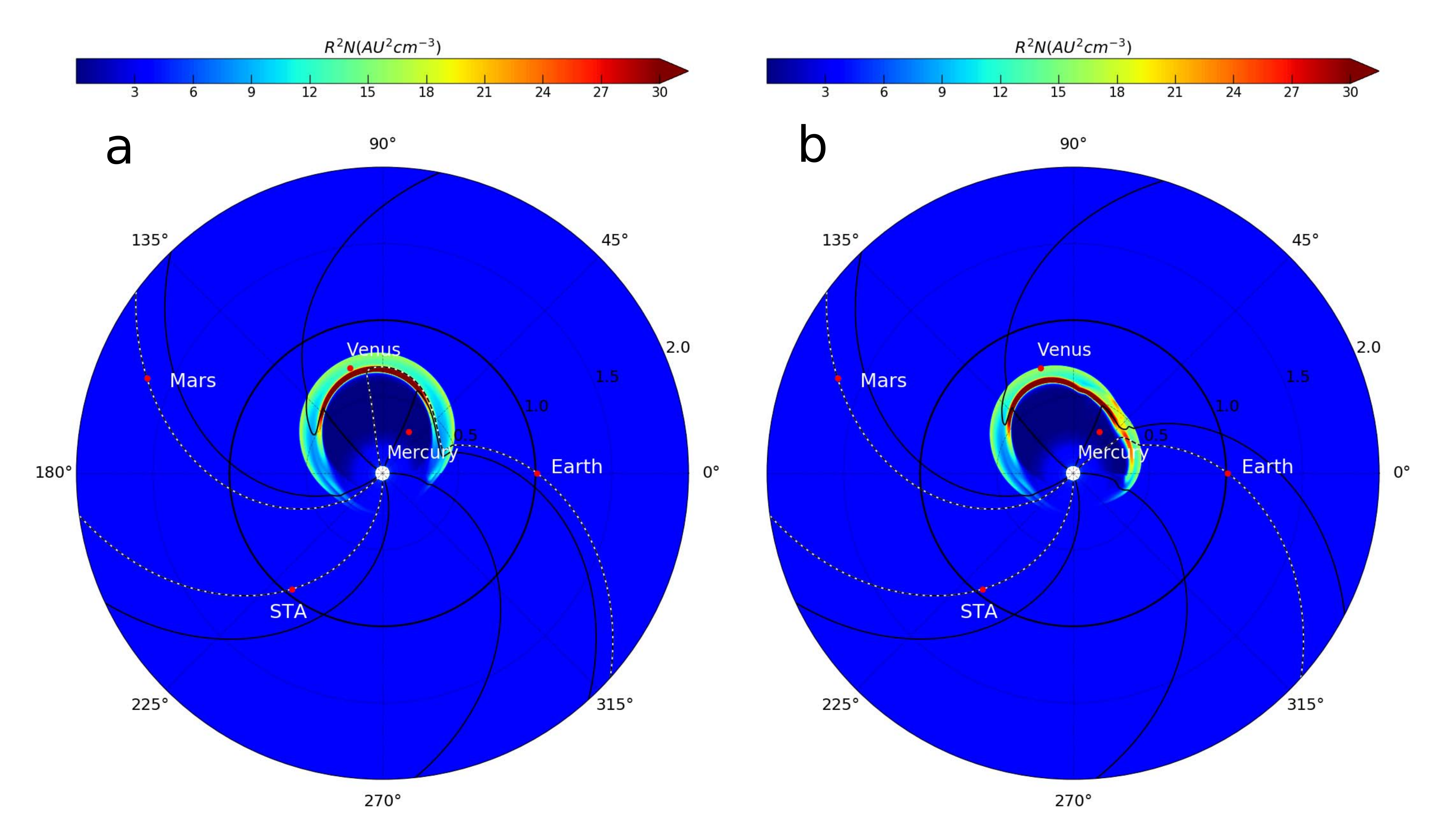}
    \caption{Snapshots of the CME-driven shock configuration. The color scheme is for the normalized density $nr^{2}$. The simulation domain is from $0.05$ to $2$ AU. The bold black circle is $r=1$ AU.  Planets Mercury, Venus, Earth and Mars and the spacecraft STEREO-A are marked as red dots, with the corresponding dashed lines the field lines passing them. The left panel shows a symmetric eruption with a single Gaussian velocity profile and the right panel shows the lateral expansion of CME-driven shock with a two-Gaussian profiles. }
    \label{fig.snapshot}
\end{figure}

Figure~\ref{fig.snapshot} shows the location of various spacecraft when the event occurred and plots the scaled number density for two cases: one for the single Gaussian perturbation and the other for a double Gaussian perturbation. When the CME erupted, Earth was located at $r=1.0$ AU with  $\phi=0^\circ$; and Mars was located at $1.66$ AU and $\phi=157.5^\circ$, STEREO-A is located at $0.96$ AU and  $\phi=232^\circ$. Venus and Mercury are on the propagation path of this CME. However, there were no available data at these two locations. The CME-driven shock arrived at Earth and Mars at $\sim$50.5 hours and $\sim$59 hours, respectively, after the eruption (\citealt{Guo+etal+2018}).
Previous work by \citet{Guo+etal+2018} of this event has obtained a speed of $2600$ km s$^{-1}$ at $17.6 R_{s}$ with the central axis at 110$^{\circ}$. In another study by \citet{Luhmann+etal+2018} employing the Cone model,  the authors obtained a radial speed of $2500$ km s$^{-1}$ at $21.5 R_{s}$ with the center axis at 108$^{\circ}$. In our simulation with a single Gaussian profile, shown in Figure~\ref{fig.snapshot} (a),
we assumed the central axis is along  $\phi_{c}$= $100^{\circ}$ and the variance $\sigma$ of the Gaussian disturbance was set to be $42^\circ$.
At the inner boundary located at 0.05 AU ($\sim 10R_{s}$) we increase the solar wind speed to $2400$ km s$^{-1}$  and
the number density from the background by a factor of $4$ at $\phi_{c}$, and the disturbance lasts $1$ hour.
With these parameters, our simulated shock arrival time is $49.2$ hours at Earth and  $58.5$ hours at  $\phi=155^\circ$ near Mars, in reasonable agreement to
observations.  However, with these parameters, the  eastern flank of the shock, magnetically connecting to Earth, is very weak and it does not accelerate particles to high energies. As noted by \citet{Liu+etal+2019}, large lateral expansion of CME-driven shock could lead to enhanced SEPs for observers that are connected to the shock flank.
To evaluate the effect of the large lateral expansion, panel (b) of Figure~\ref{fig.snapshot} shows the shock profile using a  double-Gaussian profile with parameters given in Table~\ref{table1}. As we can see, a high density blob now appears in the eastern flank, leading to a locally enhanced shock compression ratio. The modelled shock arrival times are also similar with observations: $48.5$ hours of Earth and $58.1$ hours in the longitude of 155$^\circ$ near Mars. The shock speed at Earth is about $570$ km s$^{-1}$  (\citealt{Liu+etal+2019}), and the modelled shock speed at Earth is about $650$ km s$^{-1}$.

\begin{figure}[htb]
    \centering
    \noindent \includegraphics[width=1.0\textwidth]{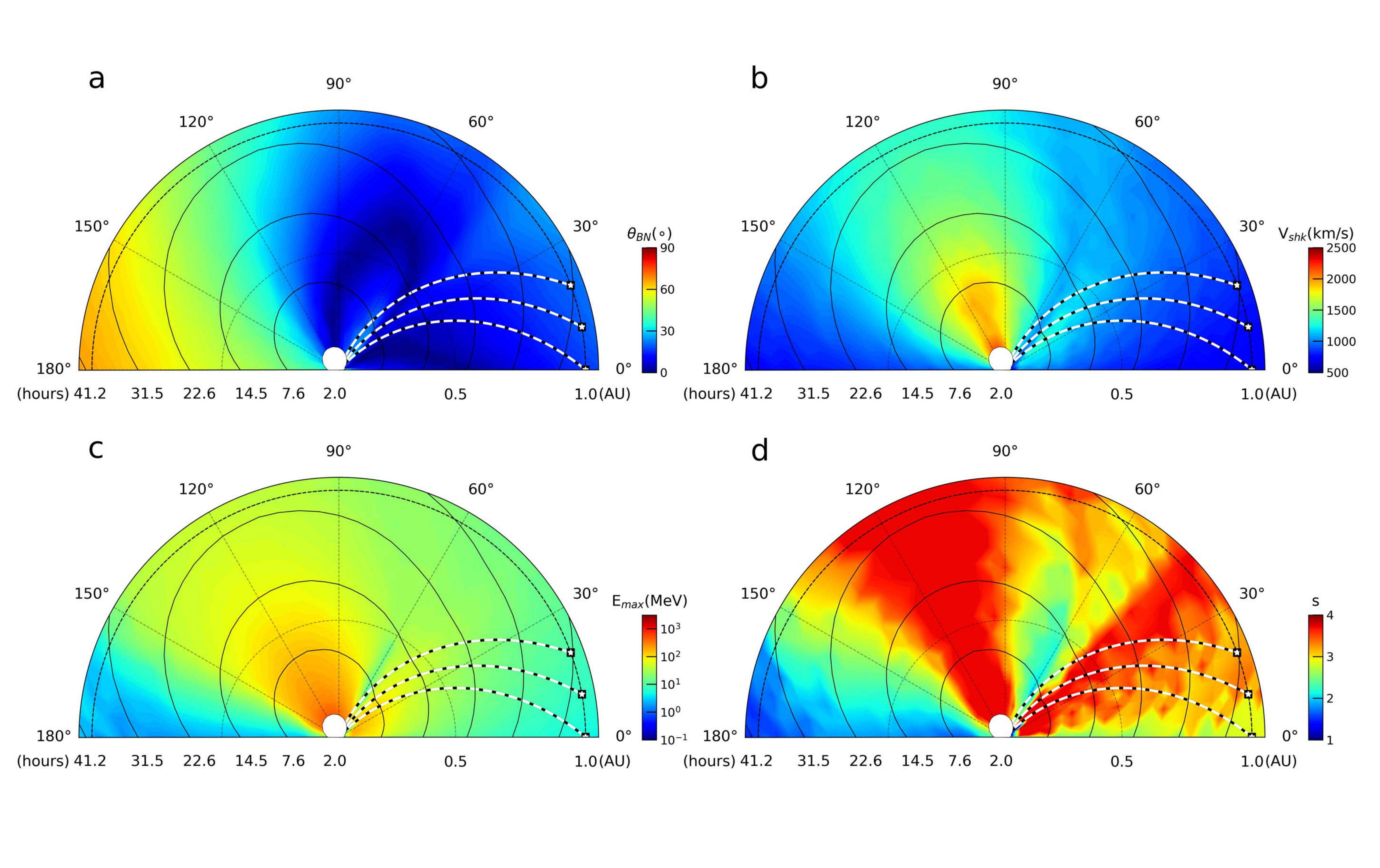}
    \caption{The time evolution of the shock parameters along the shock front. The black solid curves indicate shock fronts at different times (labelled on the left part of the x-axis). The white dash lines represent the unperturbed Parker magnetic field lines connecting observers at longitudes of 0$^\circ$, 10$^\circ$, 20$^\circ$.  Earth is located at $\phi=0^\circ$. The
    color schemes for panel (a) is the shock obliquity angle $\theta_{BN}$ (the angle between shock normal and the upstream magnetic field); for panel (b) the shock speed; for panel (c) the maximum particle energy in the shock front; and for panel (d) the shock compression ratio.}
    \label{fig.fig2_shk}
\end{figure}

Figure~\ref{fig.fig2_shk} plots the evolution of shock parameters in the ecliptic plane. Three observers at 1 AU with different longitudes: 0$^\circ$,10$^\circ$,20$^\circ$ are chosen. The white dash lines in the figure are interplanetary magnetic field lines assuming a solar wind speed of $500$ km/s. From panel (a) we can see that the eastern flank of the shock maintains a quasi-parallel geometry for the entire propagation, implying that the eastern flank of the shock has a higher efficiency to accelerate particles. Panel (b) shows the evolution of the shock speed.  We note that the shock speed is smaller than $1200$ km/s most of the time when Earth connected to the shock fronts. Panel (c) shows the maximum particle energy at shock fronts as the shock propagates from $10 R_s$ to $1$ AU.  The maximum energy occurs near the Sun at a longitude $\sim 110^{\circ}$.
However, even along the field line connecting to Earth, the maximum energy can reach hundreds MeV early on.This is due to large lateral expansion: the presence of the second Gaussian profile leads to a larger compression ratio and a larger shock speed in the eastern flank.
Panel (d) shows the compression ratio along the shock front during the shock propagation.
There are two high-compression-ratio lobes, corresponding to the two velocity profiles. The high compression ratio in the eastern flank enhances the acceleration efficiency, indicating the importance of the large lateral expansion.

\begin{figure}[htb]
    \centering
    \noindent \includegraphics[width=1.0\textwidth]{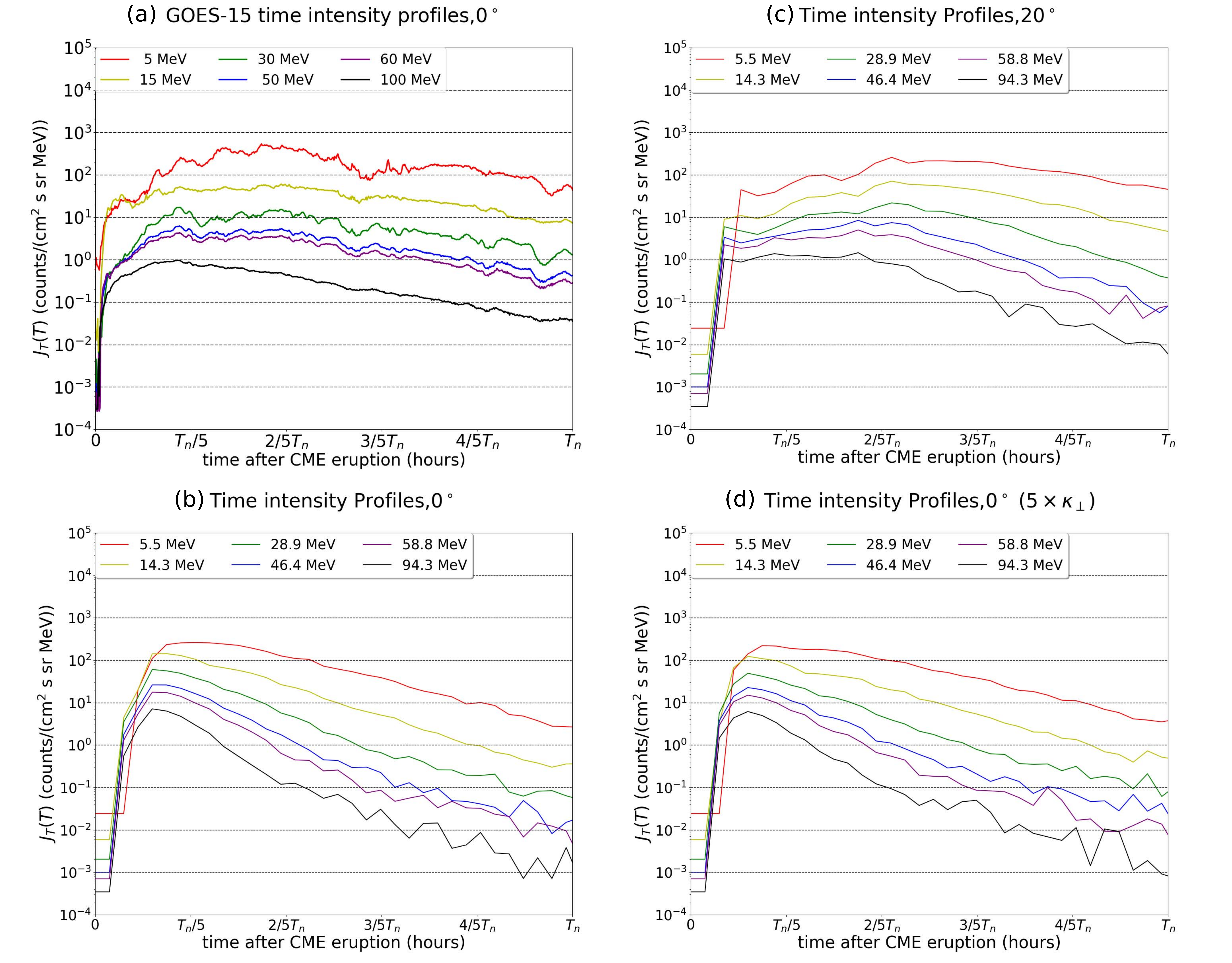}
    \caption{ panel (a) shows the proton time intensity profile measured by the GOES-15
    from onset to shock arrival. Other panels show the simulation results before shock arrival at different longitudes ( 0$^\circ$, 20$^\circ$). The curves of different colors correspond to different particle energies. The x-axis of all panels are the time since the eruption, normalized to $T_{n}$, where $T_{n}$ is the shock arrival time for different observers.}
    \label{fig.fig3_time_intens}
\end{figure}

In Figure~\ref{fig.fig3_time_intens},  the modelled time intensity profiles, shown in panels (b) to (d), at longitudes of $0^\circ$ and $20^\circ$ are compared with the observations of the Energetic Proton, Electron, and Alpha Detector (EPEAD-A) instrument on board GOES-15, shown in panel (a). Note that the EPEAD-A detector has a westward-viewing angle. From our model, we choose energy channels that are similar to the observations.
Because the shock arrival time differs for different longitudes, the $x$ axis is normalized by $T_{n}$,  the shock arrival time for each observer. From panel (a) we can see that the observed intensity at all energy channels show substantial enhancements at the beginning of the event and  are followed by slow decays.
Panel (b) presents the simulated result at Earth (i.e. $\phi=0$). The intensity shows a rapid enhancement at the beginning, similar to the observation. However, the decay is also very rapid, which differ considerably from the observation. As shown in panel (b) and (d) of Figure~\ref{fig.fig2_shk}, this is due to the fact that after $\sim 0.5$ AU, Earth is connected to a portion of the shock that has a weaker compression ratio and a low speed.  Note that this connection is under the assumption that the IMF is given by a Parker field. There was some preceding CMEs in this event and the magnetic connection can therefore be different from the Parker field. We will discuss this later.
Panel (c) is the simulated result at longitude of 20$^\circ$. It is interesting to note that simulation results at longitudes 20$^\circ$ show similar decay behavior as the observation. This observer maintains relatively long duration connecting to regions of the shock that have high compression ratios. The presence of this long-duration high compression ratio is the result of the second Gaussian profile.  Clearly, the assumption of the large lateral expansion of CME-driven shock is crucial for our results. We also note that if the preceding CMEs perturb the IMF such that Earth is also magnetically connected to the second Gaussian for an extended period, then we would expect observers at Earth could see results similar to panel (c).  Panel (d) shows the simulated result at Earth with five times as large as $\kappa_\perp$. The results between panel (b) and panel (d) are only slightly different, which illustrates that by invoking a large perpendicular diffusion alone can not explain the Earth observation.
One may think that by aligning the center axis of the second Gaussian profile closer to the Earth direction can improve the simulation results at
$\phi=0^{\circ}$ with an ambient field that is Parker like. However, the profile of the second Gaussian profile is constrained by the shock arrival time at Earth.  If one further tunes the center axis of the second Gaussian profile toward $\phi=0^{\circ}$, the shock arrival time will be
earlier than the observation. As a consequence, one has to decrease the amplitude of the second Gaussian,  which will lead to a weaker shock and
does not accelerate particles to high enough energy early on. Furthermore, if the second Gaussian profile is further towards Earth, the overall CME and shock profile will be quite distorted and appear more like two separate eruptions, which is not supported by  observations.

\begin{figure}[htb]
    \centering
    \noindent \includegraphics[width=1.0\textwidth]{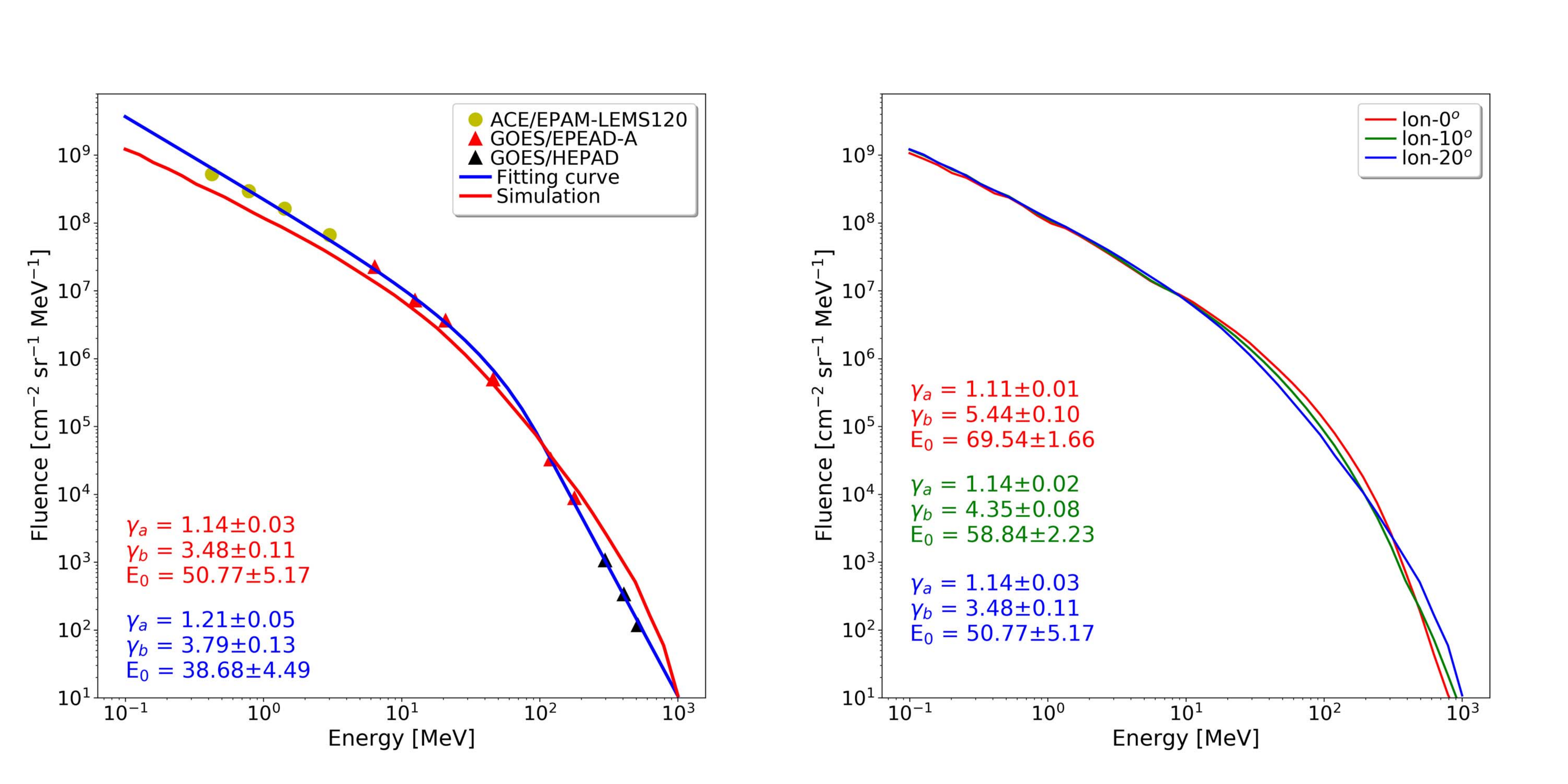}
    \caption{Time interval-integrated proton fluence. Left panel: The yellow solid dots represent the fluences from the instrument EPAM-LEMS120 on board the ACE. The red solid triangle dots and black solid triangle dots represent the fluences with the instrument EPEAD-A and HEPAD on board the GOES-15. The blue curve is the fit to the observation using the band function. The fitting parameters are shown in blue. The red curve is the modelled fluence at longitude of 20$^\circ$ with the red parameters from the band function fitting. Right panel: The modelled fluences at three different longitudes (0$^\circ$, 10$^\circ$, 20$^\circ$) and the fitting parameters are labelled with corresponding colors.}
    \label{fig.fig4_spec}
\end{figure}

Besides the time intensity profiles for the $6$ energies, we also consider the event-integrated spectra at $\phi=20^{\circ}$. Reasonable agreement with the observation is also obtained.
Left panel of Figure~\ref{fig.fig4_spec} shows the simulated
and observed event-integrated proton fluence.
For the observation, we use the lower-energy data from the Low Energy Magnetic Spectrometer-120 (LEMS-120) of the Electron, Proton and Alpha Monitor (EPAM) on board ACE, for four energy channels ranging from 330 keV to 4.75 MeV; and  the high-energy data from  EPEAD-A (P2-P6) and the High Energy Proton and Alpha Detector  (HEPAD, P8-P10) on board GOES-15. The integration period is from 16:05 UT on 10 September 2017  to 18:30 UT on 12 September 2017.

We fit both the observed and the modelled spectra with a Band-function form (\citealt{Band+etal+1993}), given by,
\beq
  \begin{split}
 &J(E) = CE^{-\gamma _{a}} \exp \left ( -\frac{E}{E_{0}} \right )  \ for\  E\leqslant (\gamma _{b}-\gamma _{a})E_{0}; \\
&J(E) = CE^{-\gamma _{a}}\left [ (\gamma _{b}-\gamma _{a})E_{0} \right ]^{\gamma _{b}-\gamma _{a}}\exp(\gamma _{a}-\gamma _{b}) \ for\  E \geqslant (\gamma _{b}-\gamma _{a})E_{0},
  \end{split}
\label{eq:band}
\eeq
where $J(E)$ is the particle fluence, $C$ is a normalization constant. $\gamma_{a}$ and $\gamma_{b}$ are the spectral indices in the low energy region and high energy region respectively. $E_{0}$ is the spectral break energy, which typically occurs at energies of tens MeV.
Following \citet{Bruno+etal+2019}, we use the calibration schemes by \citet{Sandberg+etal+2014} and \citet{Bruno+2017}, below and above 80 MeV, respectively, to obtain the mean energy values of every channel.  Using different satellite instruments, calibration schemes and integrated periods lead to different fitted parameters. \citet{Cohen+Mewaldt+2018} obtained $\gamma_{a}\sim$0.73, $\gamma_{b}\sim$3.39 and $E_{0}\sim$19.1 MeV. The smaller $\gamma_{a}$ is caused by using data from the Ultra-Low Energy Isotope Spectrometer (ULEIS) instrument on board ACE. \citet{Bruno+etal+2019} obtained  $\gamma_{a}\sim$1.21, $\gamma_{b}\sim 4.04$ and $E_{0}\sim 37.0$ MeV for a longer period, from 16:05 UT on 10 September 2017 to 00:00 UT on 17 September 2017. The fitted parameters for our period in this work are $\gamma_{a}\sim$1.21, $\gamma_{b}\sim 3.79$ and $E_{0}\sim 38.68$ MeV, similar to \citet{Bruno+etal+2019}. In comparison, the fitted parameters of the modelled fluence observed at longitude of 20$^\circ$ are: $\gamma_{a}\sim 1.14$, $\gamma_{b}\sim 3.48$ and $E_{0}\sim 50.77$ MeV.  These are similar to the
 observations.
The right panel of Figure~\ref{fig.fig4_spec} shows the simulated  event-integrated proton fluence at different longitudes of 0$^{\circ}$, 10$^{\circ}$ and 20$^{\circ}$. Note that while the time intensity profiles at these three locations are quite different as shown in Figure~\ref{fig.fig3_time_intens}, the differences of the event-integrated spectra of these three observers are however, small. Nonetheless, comparing the three spectral indices $\gamma_{b}$ in the right panel shows that the high energy portion of the spectra becomes softer as the longitude decreases. This is because at lower longitudes the observer connects to parts of the shock flank with a compression ratio decreases quicker.

This GLE event is one of the only two GLE events identified in the solar cycle 24. The high energy spectral index of this GLE event was softer than many GLE events observed in cycle 23 (\citealt{Cohen+Mewaldt+2018,Gopalswamy+etal+(2018)}).  \citet{Gopalswamy+etal+(2018)} suggested that the soft spectrum may be caused by poor latitudinal and longitudinal connectivities. Because the iPATH model is a 2D code and is limited to the ecliptic plane, we cannot consider the latitudinal effects.
As to the longitudinal connection, our results suggests that the observation at Earth (with $\phi=0^{\circ}$) is very close to the simulation results at longitude of 20$^\circ$. Note that our simulation assumes the background magnetic field is given by the Parker field. Distortion of the background field is possible due to preceding CMEs. There were multiple CMEs prior to the September 10 event. A large and fast CME occurred on the 6th, whose propagation direction is 40$^\circ$ west to Earth. At Earth,  the magnetic cloud (MC) behind the shock was observed on the 8th and ends on the beginning of 9th (\citealt{Werner+etal+2019}). It is very likely that the MC is a manifestation of a flux rope structure whose two feet are anchored on the solar surface. If so, as this flux rope can further propagate out, it can affect the field that is connected to Earth on the 10th. Besides the CME on the 6th, there were also multiple smaller CMEs occurred on the 8th and 9th. The effects of these smaller CMEs on the global magnetic field configuration are also unclear. In our simulation, we assume the background field is of Parker, so we can not address the connection issue. Note that the large lateral expansion of the CME-driven shock is crucial for obtaining our results.  \citet{Liu+etal+2019} have suggested that the enormous lateral expansion of the shock has an important effect on SEPs production of this event. However, to our knowledge,  there has been no simulations addressing SEPs that consider the effect of lateral expansion of a CME and its driven shock.
A large lateral expansion of the shock can affect both the time intensity profile and the even-integrated spectra.  This is clearly illustrated in Figure~\ref{fig.fig2_shk},
from which we can see that the observer at $\phi=20^{\circ}$  can initially connect to  a weak part at the edge of first Gaussian profile (compression ratio $\sim2$) of the shock, and followed by  a stronger part (compression ratio $ \sim4$) of the shock.

\begin{figure}[htb]
    \centering
    \noindent \includegraphics[width=1\textwidth]{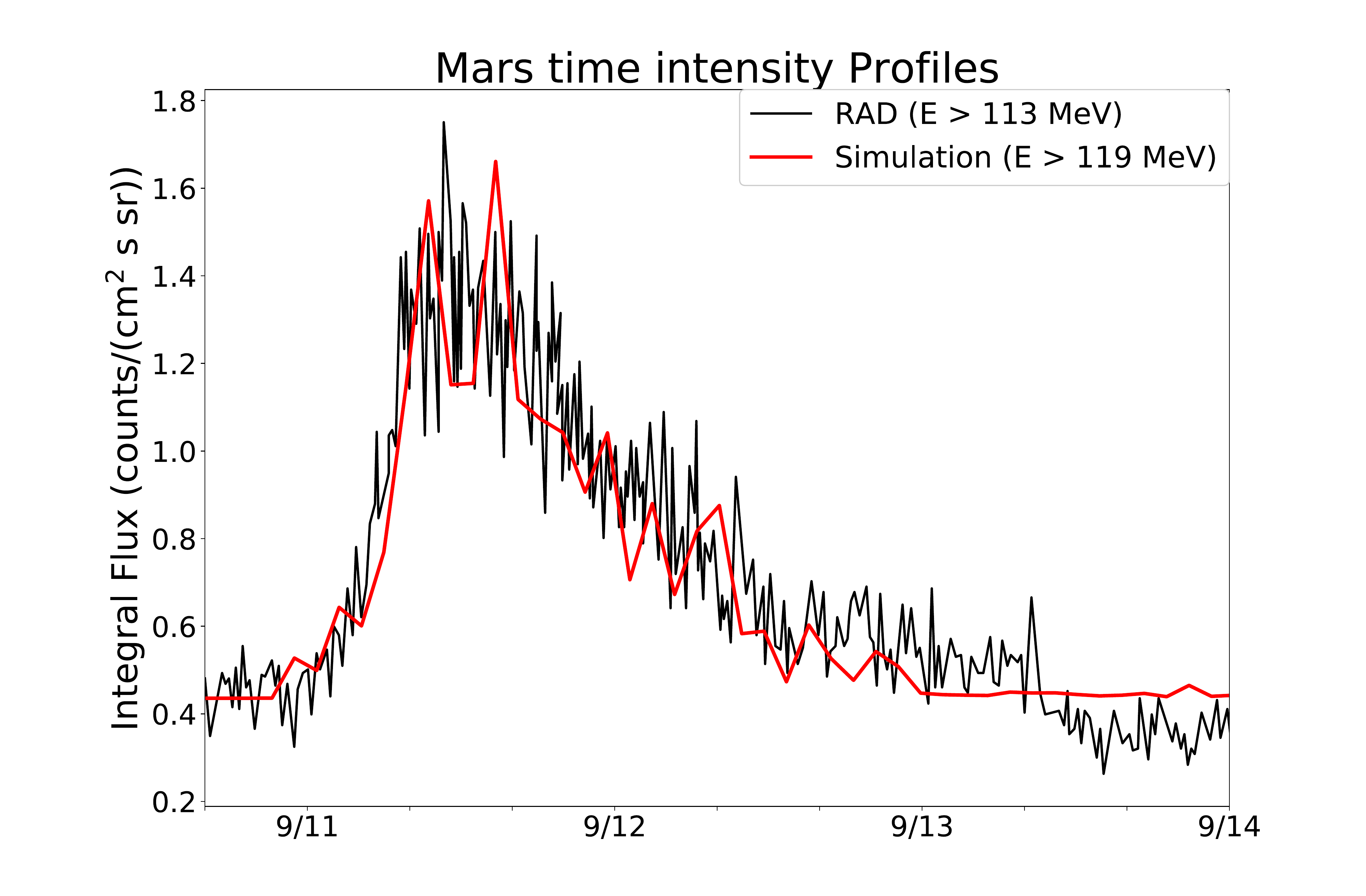}
    \caption{The proton time intensity profiles from the observation and simulation at Mars. The black curve shows the integral flux in the energy bands $>$113 MeV of the instrument RAD on the MSL (Adapted from \citealt{Zeitlin+etal+2018}). The red curve shows the modelled result in the energy bands  $>$119 MeV.}
    \label{fig.fig5_mars}
\end{figure}

The high energy SEPs associated with the fast CME were also detected by Mars and STA. \citet{Zeitlin+etal+2018} reported the radiation dose rate on the surface of Mars measured by the   Radiation Assessment Detector (RAD), mounted on the Mars Science Laboratory's (MSL's) curiosity rover. The black curve shown in Figure~\ref{fig.fig5_mars} is the derived integral flux of proton energy greater than 113 MeV observed by RAD (\citealt{Zeitlin+etal+2018}). The proton onset time is 19:50 on 10 September 2017 at Mars, which was $\sim$4 hours after the CME onset. A rapid increase is detected after the onset. The red curve in Figure~\ref{fig.fig5_mars} is the simulated integral flux of proton energy greater than $119$ MeV.
The event peaked on 11 September, from about 4:00 to 14:00 UT (\citealt{Zeitlin+etal+2018}). The modelled result shows a similar peak in the figure.  From panel (b) of Figure~\ref{fig.snapshot} we can see that Mars was  connected to the very western flank of the shock initially and as the shock propagates out, the connection becomes ever weaker. Many SEPs observed at Mars are particles that undergo cross field diffusion in the IMF. They are accelerated at the stronger part of the shock. To the observer at Mars, the event is a typical eastern event. Therefore we see a clear decay phase after the peak. The shock arrival time is $\sim$02:50 UT on 13 September 2017 followed by a clear Forbush decrease  (\citealt{Guo+etal+2018}). The simulated shock arrival time is $\sim$02:00 UT on 13 September 2017.

\begin{figure}[htb]
    \centering
    \noindent \includegraphics[width=0.8\textwidth]{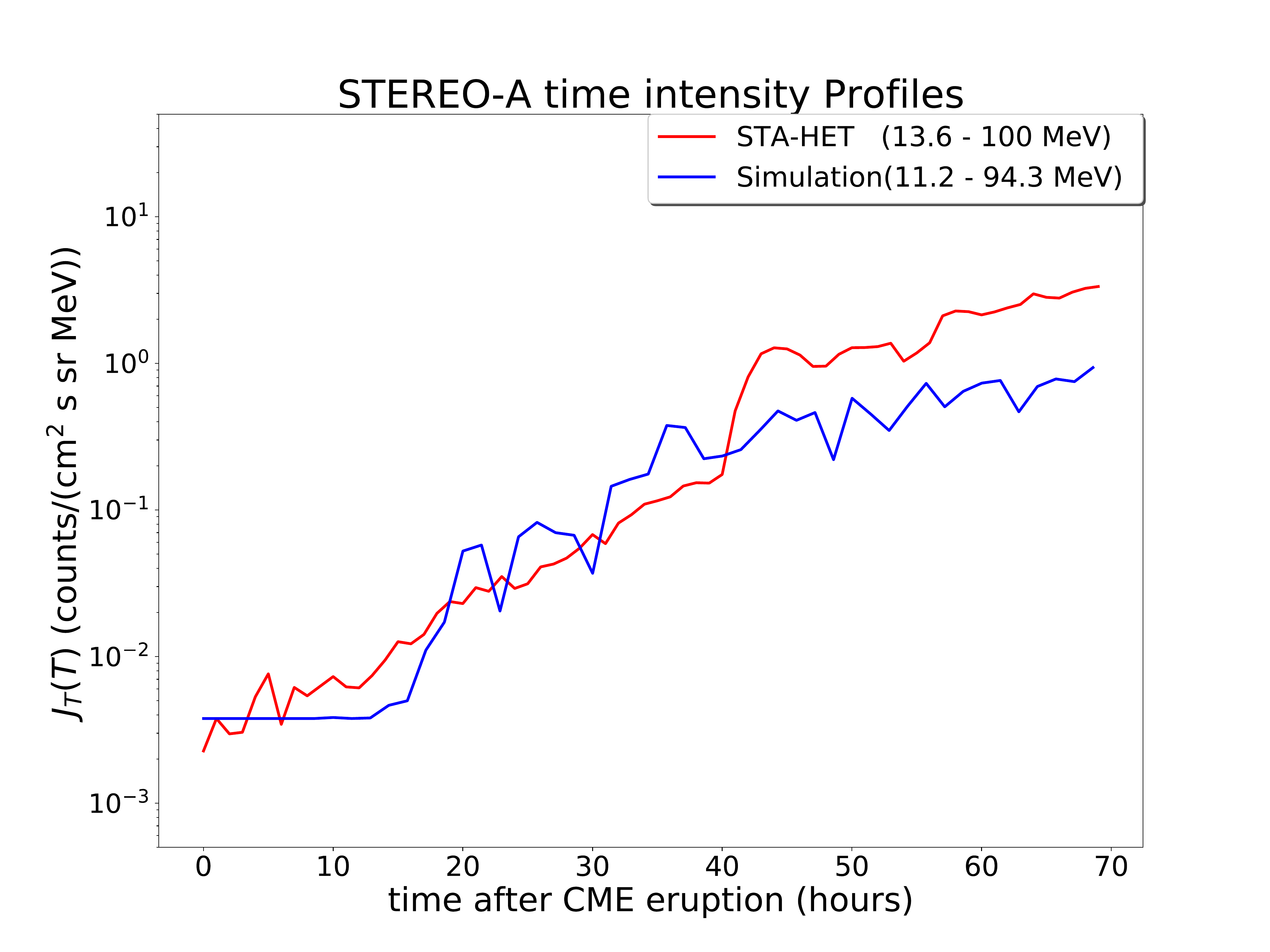}
    \caption{The proton time intensity profiles of the observation and simulation observed by STEREO-A. The red line represents the flux in the energy bands $13.6-100$ MeV of the instrument HET on board the STEREO-A. The blue line is the modelled result in the energy bands $11.2-94.3$ MeV. }
    \label{fig.fig6_sta}
\end{figure}

STEREO-A also observed this event. In Figure~\ref{fig.fig6_sta}, the modelled proton time intensity profile in the energy bands 11.2-94.3 MeV is shown as the blue curve. This is to be compared with observation of $13.6$-$100$ MeV proton (red curve) from the High Energy Telescope (HET) instruments onboard STEREO-A. Both of them show a gradual rising phase after the CME eruption.  The absolute value of the flux are also similar. The STA was at longitude of 232$^\circ$ which was connected to the back side of the sun when the CME erupted. Therefore the SEPs detected at STA must have undergone cross field diffusion as they propagate out. Our results suggest that the perpendicular diffusion coefficients and its radial dependence used in this work are reasonable.

\section{Conclusion and Discussion}
\label{sec.Conc}
In this paper, we model the 10 September 2017 SEP event using the iPATH model. The most important assumption we made in this work is a large lateral expansion of CME-driven shock, modelled by two Gaussian velocity profiles. In previous practices of iPATH modeling  (\citealt{Hu+etal+2017,Hu+etal+2018,Fu+etal+2019}), a single symmetric Gaussian-shape velocity profile is adopted to simulate the inner boundary condition of the the CME-driven shock.  To illustrate the lateral expansion, we use a velocity profile at the inner boundary consisting of two Gaussian profiles with different propagation directions and different start times. We then examine the time intensity profiles and the event-integrated spectra at multiple observation locations corresponding to observers at Earth, Mars and the STEREO-A spacecraft.
We adjust the initial CME perturbation parameters such that the modelled shock arrival times at Earth and Mars are close to the observations. The introduction of a second velocity profile to mimic the lateral expansion leads to stronger particle acceleration at the eastern flank of the shock, which is necessary to understand the observation at Earth.
Our model results at $\phi=0^{\circ}$, which is the Earth's location, show similar event-integrated spectra as the ACE and GOES observations, but the modelled time intensity profile shows a clear differences from that of observation.
In comparison, the model results at $\phi=20^{\circ}$ are comparable to ACE and GOES observations for both the time intensity profiles and the event-integrated spectra. We interpret this by a possible field line distortion due to preceding events. There were multiple smaller CMEs before this event, which can cause the IMF to be deviated from the nominal Parker configuration.
 We do point out that in the acceleration module of our model a Parker-like background IMF is assumed. When the unperturbed field line is non-Parker, it will affect the shock obliquity, and therefore the acceleration process. Modelling a CME with a non-Parker field is out of the scope of this work and will be pursued in a future work.

We also obtain time intensity profiles at longitudes that correspond to Mars and STEREO-A. The model results at these two locations are in good agreements with the observations. Because both Mars and STA are not well connected to the CME-driven shock, SEPs observed in these locations have to undergo cross-field diffusion. The agreements between the modeling results and the observations therefore provide a confirmation of our choice of the perpendicular diffusion coefficient, and its radial dependence.

Our study suggests that modeling a realistic SEP event needs 1) an in-depth understanding of the influence of preceding CMEs, in particular the effect of a non-Parker field, and 2) taking into account of possible lateral expansion. Indeed, observations have shown that large lateral expansion of CME-driven shock is quite frequent in large eruptions, so considering the lateral expansion and capturing a realistic CME-driven shock profile, such as the shock speed and obliquity along the shock surface is also crucial for any SEP modelings.  We point out that lateral expansion are intrinsically inhomogeneous because the plasma conditions at different part of the shock flank are different. In this work, we use a second Gaussian profile to mimic the lateral expansion.  Finally, fitting observations at multiple longitudes can provide very strong constraints on the perpendicular diffusion coefficients of energetic particles.  Although our simulation is for the 10 September 2017 event, it forms a nice basis for understanding other large SEP events as well.

\begin{acknowledgements}
We acknowledge the use of NOAA GOES data
(https://satdat.ngdc.noaa.gov/), the ACE data from the ACE Science Center (www.srl.caltech.edu/ACE/ASC),  the STEREO data (https://stereo-ssc.nascom.nasa.gov/).
The work done at the University of Alabama in Huntsville are partially funded by NASA grants NNX17AI17G, 80NSSC19K0075 and 80NSSC19K0629.

\end{acknowledgements}

\bibliographystyle{raa}
%\bibliography{ms2020-0054}


\label{lastpage}

\end{document}